# Erosion of Copper Target Irradiated by Ion Beam

S. V. Polosatkin, A.V. Burdakov[*], E.S. Grishnyaev[*], S.G. Konstantinov, A.A. Shoshin[**]

*Budker Institute of Nuclear Physics SB RAS, Lavrentev av., 11, Novosibirsk, 630090, Russia,*
*s.v.polosatkin@inp.nsk.su*
*[*]Novosibirsk State Technical University, Karl Marx av., 20, Novosibirsk, 630092, Russia*
*[**]Novosibirsk State University, Pirogova st., 2, Novosibirsk, 630090, Russia*

**Abstract –** **Erosion of copper target irradiated by deuterium ion beam with ultimate fluence is studied. The target originally destined for neutron generation represents bulk copper substrate covered by 3-mm titanium layer. The target was irradiated by deuterium ion beam generated in Bayard-Alpert type ion source with energy of ions 17.5 keV/nuclear. Maximal fluence in the center of the target achieves $2.5 \times 10^{23}$ atoms/cm$^2$. Measurements of the profile of irradiated target and estimation of fluence shows that physical sputtering is a dominating process that determines the target erosion Most interesting feature is growth of mm-size tadpole-shaped structures, localized in the cracks of the surface. RFA analysis of these structures showed extremely large (up to 60%at.) carbon content.**

### 1. Introduction

Sealed neutron-generating tubes for geophysics and scientific applications are under development in the Budker Institute of Nuclear Physics. The tubes represent a miniature accelerators, in which the deuterium ions are accelerated in an ion-optical system and are dumped on the target that is under the voltage of (80-100) kV. The target is a copper disk whose surface is covered with the µm-scale layer of titanium saturated with a deuterium. Key challenge of the development of neutron tubes is providing of the tube durability that is strongly depends on manufacturing technique and operation conditions of the target. Processes arises in the target apart on simplest – heating and physical sputtering – include also radiation damages of target material, hydrogen accumulation, retention, and phase transitions accompanied with formation of hydrides. These effects can dramatically force erosion of the target; a noted example of such erosion is a blistering often observed on the targets irradiated by ion beams [1]. Other structures except blisters can also form on the surface if the irradiation fluence sufficiently exceeds blistering threshold. Intrigue patterns observed on the neutron-generating target after the tube durability test are presented in the paper.

### 2. Conditions of target irradiation

Design and operation principles of the neutron tube are described in ref. [2]. The tube consists on ion source, accelerating gap, and neutron-generating target. Hot cathode ion source with electron oscillations in electrostatic field (Bayard-Alpert type ion source) are used for ion beam generation. The ions generated in the source are accelerated in the stationary electric field and impact the target. A target is copper substrate covered by 3 µm titanium layer with 10 nm molybdenum intermediate.

Main aim of discussed test of neutron tube was a study of tube durability at the ultimate beam fluence. In this test accelerating voltage was reduced to 35 kV for increasing of operation reliability. Since Bayard-Alpert ion source generates mostly $D_2^+$ ions [3], dominating fraction of ion beam has energy 17.5 keV/nuclear; a share of full-energy (35 keV) ions is about 6%.

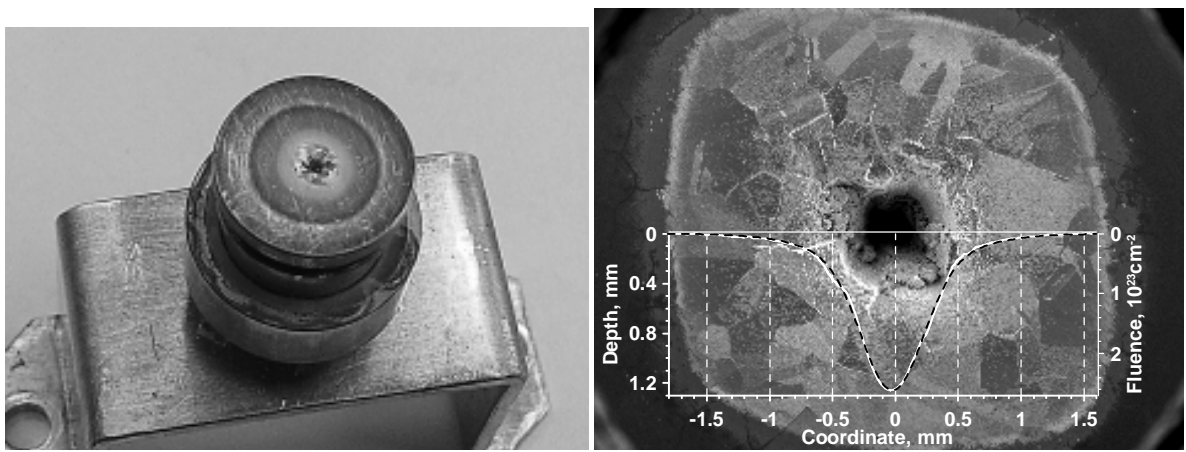

Fig.1 Photo of the target after durability test (a) and SEM image of the central part of the targets (b)

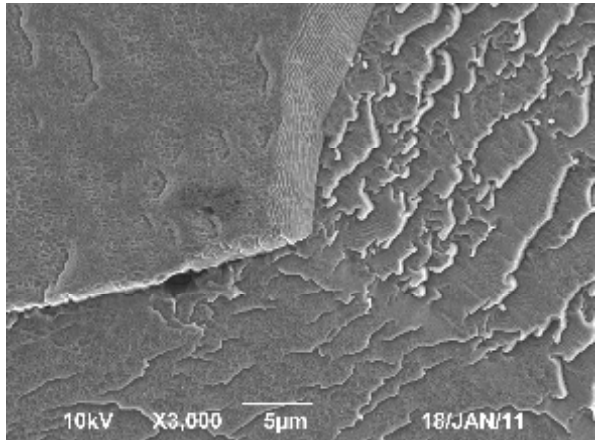

Fig.2 Flacks and open blisters on the target surface, distance 1.5 mm from the target center

A value of fluence achievable in experiment is restricted by tube operation current (50 µA) and reasonable experiment duration. For increasing of this value we focuses ion beam with a spot size on the target 0.5 mm (Fig. 1a). In the 600-operation-hours durability test the fluence in the beam spot exceeds $10^{23}$ cm$^{-2}$.

### 3. Target erosion

Several zones are observed on the surface of irradiated target. That is central hole, hole enclosing with strong erosion, and target periphery. The central hole and enclosing have diamond-shape footprint that presumably mimic grid structure of the ion source.

Profile of the target surface after irradiation is shown on Fig.1b (white solid curve). Erosion depth achieves 1.2 mm in the center of the target. The profile can be approximated by the sum of two Gaussians with characteristic diameters 0.25 and 0.52 mm and practically equal 2D integrals over target surface (black dashed curve on the Fig.2b). Possible reason for such profile is a presence of hydrogen (protium) in the ion beam. Particle tracing calculations show that difference in focusing lengths for $D_2^+$ and $DH^+$ ions can explain the observed two-component profile. Assuming linear dependence of erosion depth from a local value of fluence and taking into account full number of atoms hit to the target ($1.35 \times 10^{21}$ atoms) we can estimate spatial distribution of the fluence (right axis of Fig.2b). Maximal fluence in the center of the target achieves $2.5 \times 10^{23}$ atoms/cm$^2$. Physical sputtering under such fluence with the yield value 0.03 [4] would cause erosion depth 0.9 mm. Accordingly exactly physical sputtering is a dominating process that determines the target erosion rate in the central hole.

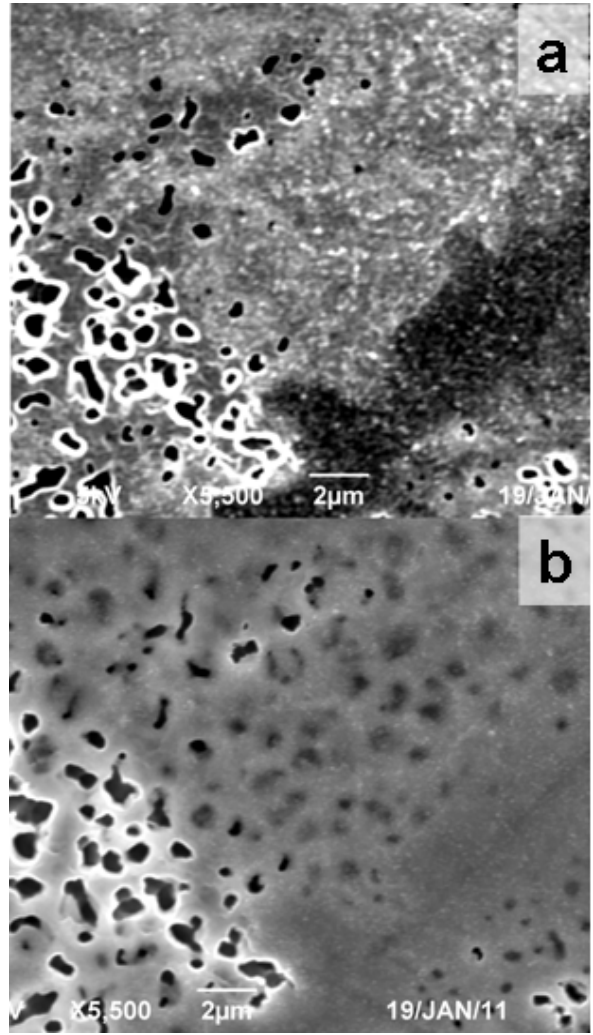

Fig.3 SEM images of intermediate zone taken with the energies of scanning electron beam 5 (top image) and 20 (bottom image) keV

Note that the estimated fluence is far above the threshold of blistering that is in order of $10^{18}$ cm$^{-2}$ [1], and signs of blisters and flacks are observed on the hole enclosure (Fig.2). Appearance of such structures strongly depends on crystal structure of the surface. Several hundreds microns grains observed in this area have sufficiently different relief of the structures on the surface.

On the target periphery a titanium layer partially preserved. A width of intermediate zone between titanium-covered and copper surfaces is about 200 µm. In this zone titanium content, measured by SEM RFA add-on, varied from 60%at. on the outer side to zero in inner side. Since RFA method gives an element content value averaged over electron range (about 3 µm for 20 keV electrons [5]), the titanium layer thickness gradually decrease in this zone from micron-scale to zero. Interesting feature of this area is appearance of cavities under surface layer. The cavities are discovered by comparison of SEM images with energies of electrons 5 and 20 keV (Fig.3). Holes in the upper right corner clearly seen on the Fig.3b ($E_e$=20 keV) and hidden on Fig.3a ($E_e$=5 keV). It follows that the holes in copper target are covered by the layer with the thickness more than 100 nm (the range of 5 keV electrons). Since the range of ions less than 100 nm that means that hydrogen accumulated in the surface layer diffuses to the bulk of target and cause cavities formation.

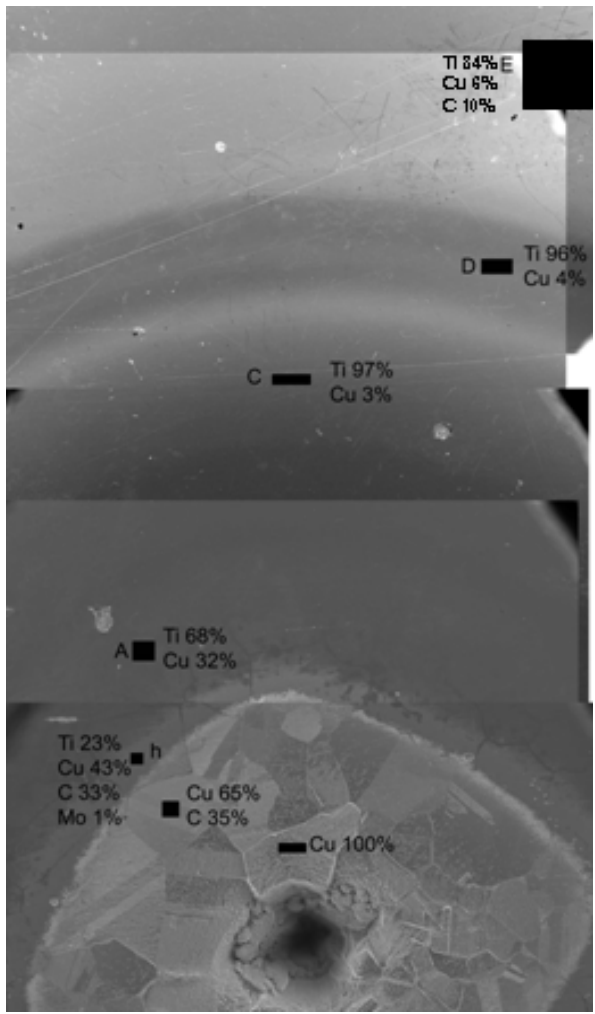

Fig.4 Elemental content of the target surface

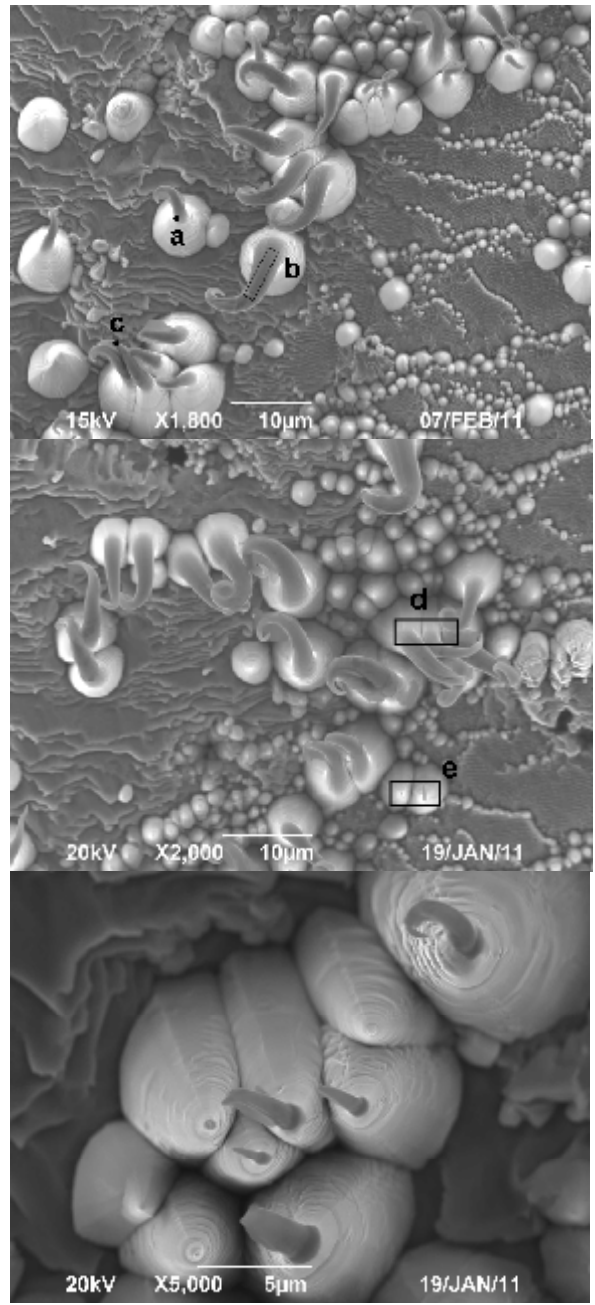

Fig.5 Freak patterns on the surface near central hole; letters marks the areas of RFA elemental content measurements, **a**- C 16% Cu 84%, **b** – C 38% Cu 62%, **c** – C 42% Cu 58%, **d** – C 60% Cu 40%, **e**- C 35% Cu 65%

**4. Carbon contamination of surface**

Elemental content of surface is measured in several points of the target by SEM RFA add-on. The target image composed from several SEM scans with denoted zones of RFA measurements and resulted elemental content is shown on Fig.5. Apart of original target elements –titanium, copper, and molybdenum the one observed impurity is carbon. According to our view origin of 10%at. carbon content on the target border is surface contamination. A surface of a zone closer to the center of the target is etched by halo of the ion beam and so clear out from carbon. On the surface irradiated by concentrated beam there are two concurrent processes – surface erosion and implantation of carbon by contaminated ion beam. Presumably balance of these processes cause accumulation of carbon in the intermediate zone and outer part of hole enclosure and clearing of surface near the central hole.

## 4. Freak patterns on the target surface

Most intrigue effect of target irradiation is growth of freak tadpole-shaped patterns (Fig.6) consisted of bell basement and spin tail. These patterns predominantly appeared on the grain borders close to central hole. Several groups of the patterns with different ages are observed on the target. Brightness of SEM images of the structures indicates good conductivity of surface. Thereat RFA analysis registers increased carbon content in the patterns. This content sufficiently varies for different patterns and achieves 40%at. for tails and 60%at. for basements with remaining copper and absence of any other impurities. At the same time RFA measurement integrated over 50x150 μm area included patterns result 100% copper content. Certainly the patterns forms by extrusion from intergranular space but specific mechanism and chemical structure of the patterns are unclear.

## Acknowledgements

This work was financially supported by the Ministry of Education and Science of the Russian Federation Government, programs "Scientific and educational staff of innovative Russia" and "National University of Science and Technology", and the Integration Interdisciplinary Project of SB RAS № 104.